%% file: main.tex
\documentclass[10pt,twocolumn,letterpaper]{article}

\usepackage[pagenumbers]{cvpr}              
\usepackage{microtype}
\usepackage{graphicx}
\usepackage{booktabs} 
\usepackage{subcaption}
\usepackage{tcolorbox}
\usepackage{mathtools}
\usepackage{amsmath}
\usepackage{amssymb}
\usepackage{mathtools}
\usepackage{amsthm}
\usepackage{chemformula}
\usepackage{makecell}
\usepackage[symbol]{footmisc}

\input{preamble}

\definecolor{cvprblue}{rgb}{0.21,0.49,0.74}
\usepackage[pagebackref,breaklinks,colorlinks,allcolors=cvprblue]{hyperref}

\def\confName{CVPR}
\def\confYear{2025}

\title{Improved Learning of Molecular Energetics Through an Electron-Wise Joint Charge Density and Energy Objective}

\author{
Vadim Ionas$^{1}$,
Jonas Elsborg$^{\ddagger,1,2}$,
Felix Ærtebjerg$^{\ddagger,1}$,
Arghya Bhowmik$^{\dagger,1,2}$\\
$^{1}$Department of Energy Conversion and Storage\\
Technical University of Denmark\\
$^{2}$CAPeX Pioneer Center\\
Kgs. Lyngby, Denmark\\
{\tt\small \texttt{arbh@dtu.dk}}\\
\\
$^{\ddagger}$These authors contributed equally\\
$^{\dagger}$Corresponding author
}

\begin{document}
\maketitle
\input{sec/0_abstract}    
\input{sec/1_intro}
\input{sec/6_multi-modal}
\input{sec/3_methods}
\input{sec/4_experiments}

\input{sec/7_conclusions}
\input{sec/8_acknowledgements}
\input{sec/9_code}
{
    \small
    \bibliographystyle{unsrtnat}
    \bibliography{main}
}


\newpage
\appendix
\renewcommand{\thetable}{\Alph{section}.\arabic{table}}
\renewcommand{\thefigure}{\Alph{section}.\arabic{figure}}
\setcounter{table}{0}
\setcounter{figure}{0}
\onecolumn
\input{sec/appendix}

\end{document}

%% file: preamble.tex
\usepackage{float}
\usepackage{booktabs}
\usepackage{placeins}


%% file: sec/0_abstract.tex
\begin{abstract}
We present a graph neural network-based learning framework trained to jointly predict electron densities and molecular energies. The model is trained to predict electron densities calculated with a Generalized Gradient Approximation (GGA) functional while simultaneously learning to predict molecular energies obtained with hybrid functional calculations at the B3LYP level of theory. Our results indicate that the rich spatial information in electron density distribution can be used to improve and accelerate the learning of accurate energies. By sharing an equivariant molecular representation across density and energy prediction heads, the model learns complementary molecular quantities within a unified framework. This provides a new promising route for practical use cases of many recent charge density learning frameworks for atomic scale simulations. 


\end{abstract}

%% file: sec/1_intro.tex
\section{Introduction}

Multi-task learning is a machine learning paradigm in which a single model is trained to predict multiple related targets through a shared representation. Rather than optimizing separate models for each quantity of interest, multi-task models use joint supervision to encourage the learned representation to capture features that are useful across tasks. In molecular modeling, this setting naturally arises when learning multiple observables of a system, either in a multi-task setting (learning different observables)\cite{queen2023polymer} or a multi-fidelity setting (learning the same observable at different levels of accuracy)\cite{ibrahim2024prediction}. The central premise is that jointly learning related tasks can improve latent representations, enhance generalization across tasks, and reduce overfitting\cite{gurnani2023polymer, kim2024data}.
However, not all combinations of learning objectives are beneficial. Since multi-task learning introduces an inductive bias through shared representations, this bias may either improve or degrade performance depending on the problem and the relationship between tasks, hence multi-task learning is most effective when the tasks are sufficiently related to benefit from shared representations while providing complementary supervision~\cite{caruana1997multitask}. In the present setting, electron density and molecular energy satisfy this requirement: the density provides spatially resolved information about the electronic structure, while the energy is a scalar quantity determined by ground state density of the molecular system.

This motivates the central hypothesis of this work: supervising a shared
equivariant representation with electron density encourages the model to
encode physically meaningful electronic-structure information, which can
improve molecular energy prediction. 

ELECTRA~\cite{electra} is an equivariant graph neural network (GNN) for predicting electron densities from atomistic structures. In the model, atoms are represented as graph nodes and neighboring atoms are connected through edges, allowing geometric information to be propagated through equivariant message passing. The resulting latent representation is used to predict the electron density while preserving the appropriate rotational transformation properties.

We test this hypothesis by extending ELECTRA from density-only prediction to joint density--energy prediction and comparing the resulting multi-task model against energy-only baselines implemented using the same underlying framework. To assess the role of density supervision, we train multi-task models using different fractions of the available density labels, corresponding to 1\(\%\), 10\(\%\), and 100\(\%\) of the density information.

%% file: sec/6_multi-modal.tex
\section{Mathematical formulation of the multi-task problem}
Density supervision can be interpreted as a representation-level regularizer for energy prediction. 
We follow the representation-learning formulation in which a predictor is written as a composition
\[
g = f \circ h,
\]
where \(h\) is a shared representation map and \(f\) is a task-specific prediction head. In our setting, the input \(x \in \mathcal X\) is a molecular structure represented as an ASE ~\cite{ase2017} \texttt{Atoms} object, containing the atomic species and Cartesian coordinates of the system. The shared representation is produced by the equivariant GNN encoder
\[
h(x) = \Phi_\theta(x) \in \mathcal Z,
\]
where \(\mathcal Z\) is the learned representation space. The energy prediction is given by
\[
E_{\mathrm{pred}}(x)
=
f_E(h(x))
=
f_E(\Phi_\theta(x)),
\]
while the density prediction is a function over real-space positions,
\[
\rho_{\mathrm{pred}}(\mathbf r; x)
=
f_{\rho}(h(x))(\mathbf r)
=
f_{\rho}(\Phi_\theta(x))(\mathbf r).
\]

For a molecule \(x\), the reference DFT quantities are denoted by \(E_{\mathrm{ref}}(x)\) and \(\rho_{\mathrm{ref}}(\mathbf r; x)\). The energy loss is the squared error
\[
\mathcal L_E(x)
=
\left(
E_{\mathrm{ref}}(x)
-
E_{\mathrm{pred}}(x)
\right)^2.
\]
The density loss is the normalized mean absolute error
\[
\mathcal L_{\rho}(x)
=
\mathrm{NMAE}(\rho_{\mathrm{pred}},\rho_{\mathrm{ref}})=
\]
\[
=\frac{
\int_{\mathbb R^3}
\left|
\rho_{\mathrm{ref}}(\mathbf r;x)
-
\rho_{\mathrm{pred}}(\mathbf r;x)
\right|
\,dV
}{
\int_{\mathbb R^3}
\left|
\rho_{\mathrm{ref}}(\mathbf r;x)
\right|
\,dV
}.
\]
Given a finite training set
\(\{(x_i,E_{\mathrm{ref},i},\rho_{\mathrm{ref},i})\}_{i=1}^{n}\),
we define the empirical risk of a predictor as the average task loss
evaluated over the training samples. Accordingly, the empirical energy
risk is the mean squared energy error over the dataset,
\[
\widehat R_E(f_E \circ h)
=
\frac{1}{n}
\sum_{i=1}^{n}
\left(
E_{\mathrm{ref},i}
-
E_{\mathrm{pred},i}
\right)^2,
\]
while the empirical density risk is the mean normalized absolute density
error over the dataset,
\[
\widehat R_{\rho}(f_{\rho} \circ h)
=
\frac{1}{n}
\sum_{i=1}^{n}
\frac{
\int_{\mathbb R^3}
\left|
\rho_{\mathrm{ref},i}(\mathbf r)
-
\rho_{\mathrm{pred},i}(\mathbf r)
\right|
\,dV
}{
\int_{\mathbb R^3}
\left|
\rho_{\mathrm{ref},i}(\mathbf r)
\right|
\,dV
}.
\]
The corresponding population risks are
\[
R_E(f_E \circ h)
=
\mathbb E_{x \sim \mu}
\left[
\left(
E_{\mathrm{ref}}(x)
-
f_E(h(x))
\right)^2
\right],
\]
and
\[
R_{\rho}(f_{\rho} \circ h)
=
\]
\[
=\mathbb E_{x \sim \mu}
\left[
\frac{
\int_{\mathbb R^3}
\left|
\rho_{\mathrm{ref}}(\mathbf r;x)
-
f_{\rho}(h(x))(\mathbf r)
\right|
\,dV
}{
\int_{\mathbb R^3}
\left|
\rho_{\mathrm{ref}}(\mathbf r;x)
\right|
\,dV
}
\right],
\]
where \(\mu\) denotes the data-generating distribution over molecular structures and DFT-derived targets.

The energy-only objective is
\[
\mathcal L_{\mathrm{E-only}}
=
\widehat R_E(f_E \circ h),
\]
whereas the joint energy-density objective is
\[
\mathcal L_{\mathrm{multi}}
=
\alpha\widehat R_E(f_E \circ h)
+
\beta
\widehat R_{\rho}(f_{\rho} \circ h).
\]
Thus, energy-only training constrains the shared representation only through the scalar energy target, while joint training additionally constrains the same representation through the spatial electron density.

This auxiliary constraint can be described in terms of the effective representation class. Let
\[
\mathcal H
=
\{h=\Phi_\theta : \theta \in \Theta\}
\]
denote the full class of shared representations induced by the GNN architecture. Let
\[
\mathcal F_E
\]
denote the class of admissible energy heads and let
\[
\mathcal F_{\rho}
\]
denote the class of admissible density heads. The energy predictor class is then
\[
\mathcal F_E \circ \mathcal H
=
\{
f_E \circ h :
f_E \in \mathcal F_E,\,
h \in \mathcal H
\}.
\]
Density supervision restricts the learned representation toward the subset
\[
\mathcal H_\epsilon
=
\left\{
h \in \mathcal H :
\inf_{f_{\rho} \in \mathcal F_{\rho}}
R_{\rho}(f_{\rho} \circ h)
\leq
\epsilon
\right\}.
\]
This is the set of representations from which the electron density can be predicted to within density risk \(\epsilon\). Since
\[
\mathcal H_\epsilon \subseteq \mathcal H,
\]
monotone complexity measures such as empirical Rademacher complexity satisfy
\[
\operatorname{Rad}_n(\mathcal F_E \circ \mathcal H_\epsilon)
\leq
\operatorname{Rad}_n(\mathcal F_E \circ \mathcal H).
\]

\begin{table*}[t]
\centering
\small
\caption{Model configurations. The frozen-representation probe (top) holds the
architecture fixed and varies only whether the encoder received density supervision;
the benchmark (bottom) compares the multi-task model against the strongest energy-only
baselines. $\mathcal{L}_E$, $\mathcal{L}_{\rho}$ as in Eqs.~(5)--(6);
``ch.'' is the HotPP channel width.}
\label{tab:model_configurations}
\begin{tabular*}{\textwidth}{@{\extracolsep{\fill}}lllll@{}}
\toprule
Model & Objective & Density labels & GNN backbone & Energy readout \\
\midrule
\multicolumn{5}{@{}l}{\emph{Frozen-representation probe (Sec.~2):}} \\
\addlinespace[2pt]
\quad $h_E$ ($\beta = 0$)
  & $\alpha\mathcal{L}_E$
  & --
  & 750 ch., frozen
  & 3-layer MLP (new) \\
\quad $h_{E+\rho}$ ($\beta = 1$)
  & $\alpha\mathcal{L}_E + \mathcal{L}_{\rho}$
  & 100\%
  & 750 ch., frozen
  & 3-layer MLP (new) \\
\addlinespace\addlinespace[2pt]
\multicolumn{5}{@{}l}{\emph{Benchmark (Sec.~5):}} \\
\addlinespace[2pt]
\quad Energy-only, 64 ch.
  & $\mathcal{L}_E$
  & --
  & 64 ch., trained
  & UGP \\
\quad Energy-only, 100 ch.
  & $\mathcal{L}_E$
  & --
  & 100 ch., trained
  & UGP \\
\quad Multi-task
  & $10\,\mathcal{L}_E + \mathcal{L}_{\rho}$
  & 1\%, 10\%, 100\%
  & 750 ch., trained
  & Projection to 100-d, UGP \\
\bottomrule
\end{tabular*}
\end{table*}

Therefore, the density objective can be interpreted as restricting the effective representation space available to the energy predictor, provided that density-consistent representations remain sufficiently expressive for energy prediction.

This interpretation is motivated by multitask representation-learning theory, in particular the analysis of Maurer, Pontil, and Romera-Paredes~\cite{multitask2015}. They study multitask predictors of the form \(f_t \circ h\), where \(h\) is a representation shared across tasks and \(f_t\) is a task-specific predictor. In their framework, the excess multitask risk admits an upper bound whose terms can be interpreted as the cost of estimating the shared representation, the cost of estimating the task-specific predictors, and a confidence term.

Theorem~1 is not used directly in our analysis, since our setting differs from the multiple-task setting considered by Maurer, Pontil, and Romera-Paredes~\cite{multitask2015}. Here, electron density and molecular energy are two physically related electronic-structure observables associated with the same molecular geometry, rather than targets arising from independent task distributions. Although the density and energy labels are obtained from separate electronic-structure calculations, they both encode properties of the same underlying molecular system and can therefore be regarded as coupled observables for the purpose of representation learning. The corresponding complexity terms are also not tractable to evaluate for our neural architecture. Nevertheless, the theorem provides useful motivation for our experimental design. It identifies the shared representation as a central object controlling multitask generalization: if auxiliary supervision encourages the shared representation to capture structure that is useful for the target task, then the task-specific predictor can in principle be learned with a reduced statistical burden. This motivates both our comparison between energy-only and density--energy multitask training and our analysis of whether density supervision improves the energy-relevant information encoded in the shared representation. These two questions are addressed using several related but distinct model configurations, summarized in \Cref{tab:model_configurations}.

To examine whether density supervision improves the learned representation itself, we perform a frozen-representation probe using two models with identical architectures and training setups, differing only in the density-loss coefficient. One model is trained with $\beta=0$, corresponding to energy-only optimization, while the other is trained with $\beta=1$, introducing density supervision in addition to the energy objective. Importantly, this comparison uses the architecture of the multi-task implementation in both cases, thereby providing a controlled baseline in which model capacity and representation dimensionality are held fixed. The energy-only model used in the final benchmark experiments employs a different energy-prediction architecture, as these experiments use the dedicated energy-only implementation to provide a stronger task-specific baseline rather than to isolate the effect of density supervision. Further details of this distinction are given in \Cref{sec:energy_prediction}, where the multi-task projection from the HotPP representation to the lower-dimensional energy channel is described. We first train the energy-only and energy-density models for the same number of epochs and then freeze their learned geometric encoders,
\[
h_E : \mathcal X \to \mathcal Z,
\qquad
h_{E+\rho} : \mathcal X \to \mathcal Z.
\]
Here, \(h\) denotes the frozen ELECTRA ~\cite{electra} base model, which maps an input structure \(x \in \mathcal X\), consisting of atom types and atomic positions, to learned geometric features. The model first constructs an atomistic graph from the molecular geometry and then applies the equivariant encoder to produce atom-wise features. In general, the encoder output contains scalar, vector, and higher-order geometric components. For the frozen energy probe, however, we use only the scalar component, since the downstream target is a scalar energy.

Let \(s_i(x)\) denote the scalar feature vector assigned to atom \(i\) by the frozen encoder. We construct the molecular representation used for downstream energy prediction by sum-pooling these scalar atom-wise features,
\[
z(x)
=
\sum_{i=1}^{N_x} s_i(x),
\]
where \(N_x\) is the number of atoms in structure \(x\). Thus, the downstream probe does not operate directly on the raw structure, atomic graph, or charge density. Instead, it receives the frozen pooled scalar representation \(z(x)\), extracted from the geometric encoder \(h\), as input.

After freezing the encoders, we discard the original energy readout heads and train identical neural-network energy probes from scratch on top of the frozen scalar representations. The resulting predictors are
\[
\tilde{E}_E(x)
=
q_{\alpha_E}(z_E(x)),
\qquad
\tilde{E}_{E+\rho}(x)
=
q_{\alpha_{E+\rho}}(z_{E+\rho}(x)),
\]
where
\[
z_E(x)
=
\sum_{i=1}^{N_x} s_{E,i}(x),
\qquad
z_{E+\rho}(x)
=
\sum_{i=1}^{N_x} s_{E+\rho,i}(x).
\]
The probe architectures and training procedures are kept identical across the two representations. Only the probe parameters are optimized, while the frozen encoders \(h_E\) and \(h_{E+\rho}\) remain fixed.

The probe parameters are trained by minimizing the empirical energy risk
\[
\alpha^\ast
=
\arg\min_{\alpha}
\frac{1}{n}
\sum_{i=1}^{n}
\left(
E_{\mathrm{ref},i}
-
q_\alpha(z(x_i))
\right)^2.
\]
This design isolates the energy-relevant information already present in the learned scalar representation, without allowing further adaptation of the base model.


In our experiment, both base models are trained for four epochs before freezing, so that the probe compares representations learned under the same training budget. After training the probes, we select the probe checkpoint with the lowest validation error and evaluate it on a held-out test set. Dataset details, train/validation/test splits, and the full experimental setup are described in Section~\ref{sec:experiments}. We emphasize that the frozen-representation probe is a diagnostic analysis and is distinct from the models used in the main benchmark experiments. The frozen representation learned under energy-only supervision achieves a test MAE of
\[
\mathrm{MAE}_{E} = 105.3~\mathrm{meV},
\]
whereas the density-supervised multitask representation achieves a test MAE of
\[
\mathrm{MAE}_{E+\rho} = 63.9~\mathrm{meV}.
\]
This corresponds to an absolute MAE reduction of \(41.4~\mathrm{meV}\), or a relative reduction of approximately \(39.3\%\). Since the probe architecture, training objective, and training procedure are identical in both cases, the lower probe error indicates that the density-supervised representation contains scalar geometric features that are more useful for downstream energy prediction. This suggests that the auxiliary density task improves the energy-relevant structure of the learned representation, rather than merely changing the original energy readout.

The choice of electron density as the auxiliary target is physically motivated. In density functional theory, the ground-state energy is fundamentally linked to the electron density. Therefore, density prediction is not an arbitrary auxiliary task: it encourages the model to encode electronic-structure information that is directly relevant to the DFT energy. From this perspective, density supervision provides an aligned inductive bias, constraining the equivariant representation toward physically meaningful features and potentially improving energy generalization. We discuss the density--energy relationship in DFT in more detail in Section~\ref{sec:related_work_motivation}.

%% file: sec/3_methods.tex
\section{Related Work \& Motivation}
\label{sec:related_work_motivation}

\paragraph{Molecular property prediction.}
Predicting properties of new molecules and materials has many applications from drug discovery to material design. The most widely used method to predict molecular properties is KS DFT due to its good balance of computational efficiency and accuracy. The central quantity in DFT is the electron or charge density $\rho: \mathbb{R}^3 \rightarrow \mathbb{R}$ defined as 
$$
\rho(\mathbf{r}) =N_e \int |\psi(\mathbf{r_1},...,\mathbf{r_{N_e}})|^2 d\mathbf{r_2}...d\mathbf{r_{N_e}}
$$
and a functional $E[\rho]$ that maps the electron density to the ground state total energy of the system. In DFT, the ground-state energy is a functional of the electron density.
The first Hohenberg--Kohn theorem establishes a one-to-one correspondence,
up to an additive constant, between the ground-state density $\rho_0$ and the
external potential $v_{\mathrm{ext}}$, implying that all ground-state
observables are, in principle, functionals of $\rho_0$ \cite{sun2019density}.
The second Hohenberg--Kohn theorem provides the corresponding variational
principle: for any $N_e$-representable density $\rho$,
\[
E_{\mathrm{HK}}[\rho] \geq E_{\mathrm{HK}}[\rho_0] = E_0.
\]
In Kohn--Sham DFT, the interacting many-electron problem is reformulated in
terms of an auxiliary noninteracting system constrained to reproduce the same
ground-state density. The total energy functional is conventionally decomposed
as
\[
E[\rho] =
T_s[\rho]
+
\int v_{\mathrm{ext}}(\mathbf r)\rho(\mathbf r)d\mathbf r
+
E_{\mathrm H}[\rho]
+
E_{\mathrm{xc}}[\rho],
\]
where $T_s[\rho]$ is the noninteracting kinetic energy, $E_{\mathrm H}[\rho]$
is the classical Hartree energy, and $E_{\mathrm{xc}}[\rho]$ contains the
exchange--correlation contribution, including the residual many-body effects
and the correction from the noninteracting to the interacting kinetic energy.
The exact form of $E_{\mathrm{xc}}[\rho]$ is unknown, making its approximation
the central challenge in practical KS-DFT. The resulting Kohn--Sham equations
are solved self-consistently to obtain an approximate ground-state density and
energy. However, conventional KS-DFT is computationally demanding for large
systems, with typical implementations exhibiting cubic scaling with system size,
primarily due to the solution of the Kohn--Sham orbitals. Orbital-free DFT
avoids the explicit orbital representation and can therefore achieve more
favorable scaling, but requires accurate density-only approximations not only
to $E_{\mathrm{xc}}[\rho]$ but also to the noninteracting kinetic-energy
functional $T_s[\rho]$, whose construction remains a major limitation
\cite{mi2023orbital}.

\paragraph{Neural network potentials.} To tackle prohibitive scaling of DFT, machine learning surrogate models trained to reproduce total energies for atomic structures at DFT accuracy but at much lower costs have gained much attention\cite{unke2021machine}. Popular high-accuracy architectures often represent the molecule as a geometric graph of atoms and use message-passing graph neural networks to directly predict energies and forces\cite{reiser2022graph} while scaling linearly with the system size due to a message passing cutoff radius. These networks can be categorized into invariant and equivariant types based on their handling of geometric information. These models have been foundational in predicting molecular properties such as energies by learning from atomic environments in a way that respects molecular symmetries. Invariant GNNs, such as SchNet~\cite{schutt2017schnet}, use scalar geometric features, such as interatomic distances, that remain unchanged under rotations and translations. A further development is provided by equivariant GNNs, which incorporate geometric features that transform predictably under symmetry operations. These models propagate not only scalar features, but also vector- and tensor-valued features constructed from local atomic geometry, allowing directional and angular information to be represented explicitly. Although molecular energies are invariant scalar quantities, equivariant models can use higher-order geometric features internally and combine them through invariant readout functions to produce scalar energy predictions. This richer representation of atomic environments has been shown to improve accuracy and data efficiency in molecular property and interatomic potential learning. NequIP\cite{batzner20223} uses E(3)-equivariant spherical convolutions, relying on spherical projection derived tensors that interact through tensor products, providing a more expressive representation of atomic interactions than invariant convolutions. This approach has shown state-of-the-art accuracy in molecular dynamics simulations, outperforming invariant models. MACE\cite{batatia2022mace} uses higher-order equivariant message passing, which enhances its ability to capture complex interatomic interactions. eSCN\cite{passaro2023reducing} improves the efficiency of equivariant convolutions by reducing SO(3) tensor products to SO(2) operations, lowering the computational cost and enabling the practical use of higher-order equivariant features. In addition to models based on spherical representations, O(3)-equivariant Cartesian tensor networks such as Tensornet\cite{simeon2024tensornet} have also been proposed. This network's equivariance is achieved by operating on Cartesian tensor features, in which feature mixing is simplified through matrix product operations, allowing the model to achieve competitive accuracy with fewer parameters and lower computational cost. Early versions could only represent tensors of order 2, but recently architectures like the High-order Tensor Passing Potential (HotPP)\cite{hotpp} have been proposed, which can handle high-order tensors, achieving expressivity and accuracy that rivals state-of-the-art spherical projection based GNNs. 

\paragraph{Electron density learning.} An alternative approach to direct energy and force predictions is to learn the electron density $\rho$, which can then be used to derive molecular properties within a physics based framework. Learning the electron density offers two potential advantages: (1) multiple molecular properties can, in principle, be derived from $(\rho)$ without requiring explicit labels for each target property, and (2) the electron density provides a spatially resolved, information-rich supervision signal that may improve data efficiency.
We can represent $\rho$ for example by evaluating a GNN at special query points corresponding to the density grid\cite{jorgensen2022equivariant}, a linear combination of atom-centered orbital basis functions \cite{fu2024recipe,febrer2024graph2mat}, or using floating Gaussian orbitals \cite{electra, elsborg2026global}. These models are similar to orbital-free DFT in that they scale linearly. Compared to NNPs they are much slower, since evaluating $\rho$ on a grid is more expensive than direct property prediction, despite their similar theoretical linear scaling. Recent work, however, has shown fast charge density inference time for charge density models like ELECTRA \citep{electra}, ELECTRAFI \citep{elsborg2026global} and BOA \citep{Klockow_boa}. Such recent developments have made multi-task models that operate on charge density and energies now computationally worthwhile to explore.

\section{Method}

As the foundation to our multi-task model, we use the charge density prediction model ELECTRA~\cite{electra}. ELECTRA is a symmetry-breaking Cartesian tensor message-passing architecture based on the HotPP Cartesian tensor network architecture~\cite{hotpp}.

The model constructs 3D charge densities in a way similar to Gaussian splatting~\cite{kerbl20233d}, using a 3D Gaussian mixture model:
\begin{equation}
    \rho\left(\mathbf{r}\right)= \sum_{A \in M} \sum_{j=0}^{N_A}w_{A,j} \mathcal{N}(\boldsymbol{r}|\boldsymbol{\mu}_{A,j},\boldsymbol{\Sigma}_{A, j}) \label{eq:ansat}
\end{equation}
where $A \in M$ represents the atoms in the molecule, and $N_A$ is the number of Gaussians for each atom, which depend on the atom type. ELECTRA models use an equivariant symmetry-breaking mechanism to model the charge density using "floating" orbitals \cite{tao1992mo, tao1993use, tasi2007hartree}, i.e. orbitals that are not centered on atoms. For additional details on the ELECTRA architecture, please refer to Ref.~\cite{electra}.

\begin{figure}[ht!]
    \centering
    \includegraphics[width=0.7\linewidth]{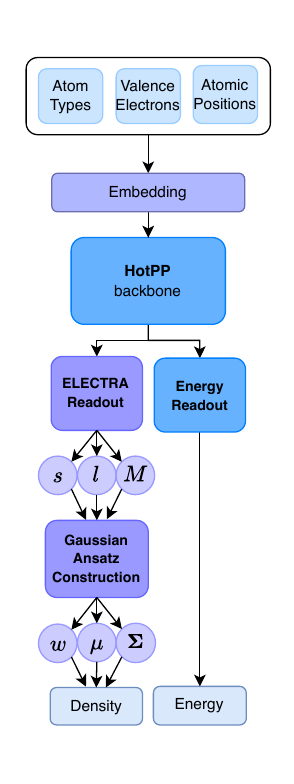}
    \caption{Overview of the ELECTRA energy–density model. Embedded atomic inputs are processed by a shared HotPP backbone into a shared representation and split into two readout branches: one predicts energy, while the other constructs the charge density from Gaussian ansatz parameters. }
    \label{fig:density_construction}
\end{figure}

\paragraph{Atomic Embedding \& Equivariant Backbone.}
The atomic structure is embedded using ELECTRA's embedding functions \cite{electra} and updated using HotPP's message passing layers to update the atomic features. The resulting embeddings thus form the shared representation that the density and energy prediction draw upon. As shown in ~\Cref{fig:density_construction}, the model uses the joint embeddings and splits into the charge density and energy prediction heads that do not overlap.


\paragraph{Variable Basis Set \& Density prediction.}
The complexity of the electronic structure generally depends on the atomic number~\cite{weigend2005balanced} as well as the number of valence electrons. ELECTRA, therefore, predicts a variable number of Gaussians depending on the number of valence electrons. This is achieved by assigning each output channel of each atom in HotPP to one Gaussian. Denoting $M_{e}$ as the number of Gaussians per valence electron, ELECTRA uses the first $N_{e} \cdot M_{e}$ channels of each atom to represent the Gaussians, where $N_{e}$ is the number of valence electrons for that atom. This provides inductive bias to the model capacity, focusing the computational effort into atomic regions of high density with many electrons.

ELECTRA uses three readout heads to produce three sets of features for each atom. These consist of $l=0$ ($\mathbf{s}$), $l=1$ ($\mathbf{v}$) and $l=2$ ($\mathbf{M}$) features, thus resulting in the sets $\mathcal{S}_{1}=\left(\mathbf{s_1}, \mathbf{v}_1, \mathbf{M}_1\right)_{A,j}$, $\mathcal{S}_{2}=\left(\mathbf{s_2}, \mathbf{v}_2, \mathbf{M}_2\right)_{A,j}$ and $\mathcal{S}_{3}=\left(\mathbf{s_3}, \mathbf{v}_3, \mathbf{M}_3\right)_{A,j}$, where $A$ indexes the atoms in the molecule, and $j$ the channel. Depending on the atom type, the channel index is $j\in [0,...,N_e(A) \times M_e]$, where $N_e$ is the number of valence electrons of that atom, and $M_e$ is the number of Gaussians per valence electron, which is controlled as a hyperparameter. These three sets are used to predict the weights $w_{A, j}$, mean 3D positions $\boldsymbol{\mu}_{A,j}$ and covariance matrices $\boldsymbol{\Sigma}_{A, j}$ for each atom. Since ELECTRA does not use all channels of the tensor network for each atom, the number of "occupied channels" $N_e(A) \times M_e$ for atom A are split into $N_e(A)$ nodes after the tensor network updates. The networks that convert the three sets of output to input for \cref{eq:ansat} thus treat the density prediction as a \textit{per-electron} task, rather than a \textit{per-atom} task, mirroring the underlying physics more closely. 
We illustrate the density construction in ~\Cref{fig:density_construction}.

\paragraph{Energy prediction}
\label{sec:energy_prediction}
To predict energies using ELECTRA, we construct a final scalar readout function. By using universal graph pooling (UGP) \cite{navarin2019universal} for aggregating the atomic features from HotPP, the model is given flexibility to adapt in the multi-task paradigm. UGP uses two fully-connected neural networks, $\omega$ and $\phi$, as well as a sum pooling to predict a single scalar output for a full graph. In the canonical per-atom paradigm, this takes the form:

\begin{equation}
E_{M}=\omega\left(\sum^{N_{Atoms}(M)}_{A \in M} \phi\left(s_{A}\right)\right),
\label{eq:unigraphpool_atom}
\end{equation}
where $E_{M}$ is the predicted energy variable, and $s_{A}$ is a scalar feature vector belonging to atom A. In the multi-task model, the HotPP features are first projected to a lower-dimensional energy channel before being passed to the scalar readout; the corresponding hyperparameters are listed in~\Cref{tab:model_hyperparameters}.


\paragraph{Training and objective function}
Prior to loss calculation, the densities predicted by the model are normalized to the number of valence electrons in the system:
\begin{equation}
    \quad \rho_{\text{pred}}(\mathbf{r}) = \rho(\mathbf{r}) \times \frac{n_{\mathrm{elec}}}{\int_{\mathbb{R}^3}\rho(\mathbf{r}) \, \mathrm{d}V},
    \label{eq:elec_normalization}
\end{equation}
which is then used to calculate the density loss $\mathcal{L}_{\rho}$:
\begin{equation}
    \mathcal{L}_{\rho} = \frac{\int_{\mathbb{R}^3}|\rho_{\mathrm{ref}}(\boldsymbol{r})-\rho_{\mathrm{pred}}(\boldsymbol{r})| \mathrm{d} V}{\int_{\mathbb{R}^3}|\rho_{\mathrm{ref}}(\boldsymbol{r})| \mathrm{d} V},
    \label{eq:densloss}
\end{equation}
The energy loss $\mathcal{L}_{E}$ is a squared-error loss on the predicted energy for the molecule (measured in Hartree):
\begin{equation}
    \mathcal{L}_{E} = \left(E_{ref} - E_{pred}\right)^2
    \label{eq:enloss}
\end{equation}
The total loss $\mathcal{L}$ is a sum of the two for the joint objective ({E+$\rho$}) model:
\begin{equation}
    \mathcal{L}_{\text{E}+\rho} = 
    \alpha\mathcal{L}_{E}+
    \beta\mathcal{L}_{\rho}.
\end{equation}
The weighting factor $\beta$ balances the density and energy objectives; its numerical value is given in~\Cref{tab:model_hyperparameters}. For the energy-only models, we simply use $\mathcal{L}_{E}$. In addition, the energy-only models are stripped of density-specific modules of the original ELECTRA implementation, such as the symmetry-breaking mechanism and debiasing layers~\cite{electra}.

%% file: sec/4_experiments.tex
\section{Experiments}
\label{sec:experiments}

\paragraph{Dataset and implementation.}
We use reference densities from the QM9 density files~\cite{jorgensen2022equivariant} generated in VASP~\cite{kresse1993ab} using the PBE~\cite{perdew1996generalized} functional and the Projector-Augmented Wave (PAW)~\cite{blochl1994projector} method. We pair the density data with energy predictions for the QM9 dataset that were calculated at the B3LYP/6-31G(2df,p) level of quantum chemistry~\cite{qm9_2}. We use the standard QM9 split consisting of \(123{,}835\) training molecules, \(50\) validation molecules, and \(10{,}000\) test molecules. Rather than training directly on the total molecular energies, we use formation energies obtained by subtracting the corresponding atomic reference contributions. For a molecule with total energy \(E_{\mathrm{tot}}\), the formation energy is defined as
\[
E_{\mathrm{form}}
=
E_{\mathrm{tot}}
-
\sum_{a \in \{\mathrm{H},\mathrm{C},\mathrm{N},\mathrm{O},\mathrm{F}\}}
n_a E_a^{\mathrm{ref}},
\]
where \(n_a\) denotes the number of atoms of element \(a\) in the molecule and \(E_a^{\mathrm{ref}}\) is the corresponding atomic reference energy.

For example, for a molecule with composition \(\mathrm{C}_2\mathrm{H}_6\mathrm{O}\),
\[
E_{\mathrm{form}}
=
E_{\mathrm{tot}}
-
\left(
2E_{\mathrm{C}}^{\mathrm{ref}}
+
6E_{\mathrm{H}}^{\mathrm{ref}}
+
E_{\mathrm{O}}^{\mathrm{ref}}
\right).
\]
The atomic reference energies for \(\mathrm{H}\), \(\mathrm{C}\), \(\mathrm{N}\), \(\mathrm{O}\), and \(\mathrm{F}\) were estimated from a small set of molecules, as described in ~\Cref{app:atomic_references}. We train all models until convergence and compare the learning curve and training dynamics of the multi-task and single-task architectures. During validation and testing, we use a single 3090 GPU and process each molecule's grid points sequentially in chunks, similar to other implementations~\cite{fu2024recipe}. If one is only interested in the energies, the evaluation on the grid can be omitted during inference.
\paragraph{QM9 Results and discussion.}
\begin{table*}[!t]
\centering
\caption{Test-set MAE (meV) for energy-only and multi-task models at different training-set sizes.}
\label{tab:comparison}
\begin{tabular}{lccccc}
\toprule
\textbf{Samples}
& \textbf{Energy only, 64 ch.}
& \textbf{Energy only, 100 ch.}
& \textbf{100\% density}
& \textbf{10\% density}
& \textbf{1\% density} \\
\midrule
1k   & 481.64 & 655.79 & 321.09 & 505.59 & 574.16 \\
10k  & 188.85 & 263.32 & 47.16  & 71.35 & 129.88 \\
50k  & 59.87  & 100.27 & 19.37 & 22.34 & 35.18 \\
123k & 36.46  & 53.74  & 14.48 & 14.83 & 18.75 \\
\bottomrule
\end{tabular}
\end{table*}

\begin{figure}[!t]
    \centering
    \includegraphics[width=\linewidth]{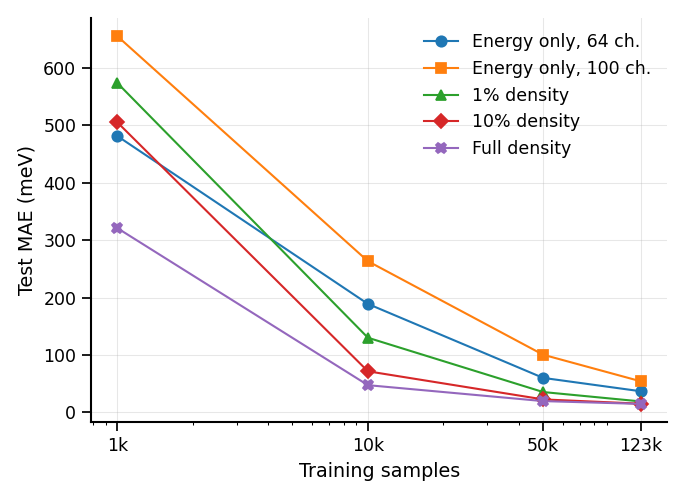}
    \caption{Test-set energy MAE as a function of training-set size for the
    energy-only and multi-task models. The multi-task model maintains lower
    MAE across all training-set sizes, indicating improved data efficiency.}
    \label{fig:learningcurves}
\end{figure}

\begin{figure}[!t]
    \centering
    \includegraphics[width=\linewidth]{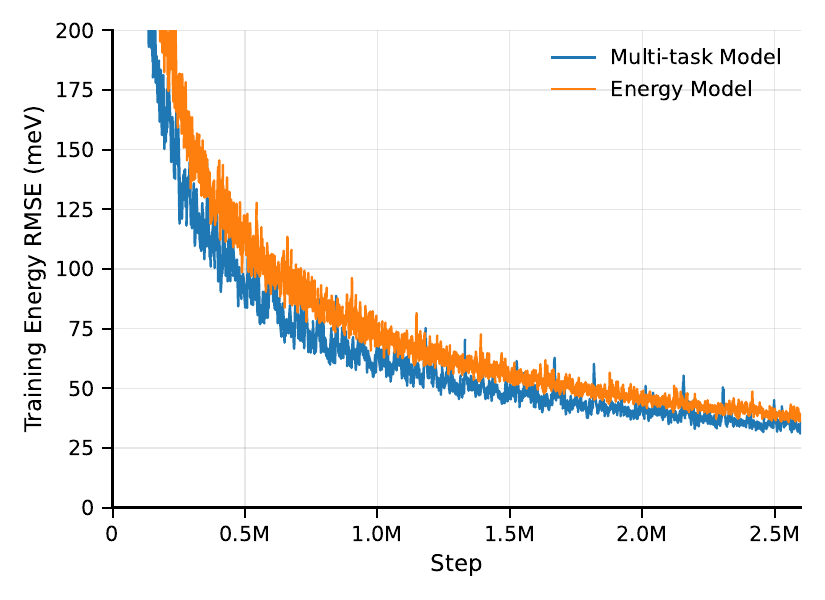}
    \caption{Training energy RMSE over twenty epochs, smoothed using a trailing
    moving-average window of 1000 steps. The multi-task model converges faster
    and reaches a lower training error than the energy-only model.}
    \label{fig:trainingcurves}
\end{figure}

\begin{figure}[!t]
    \centering
    \includegraphics[width=\linewidth]{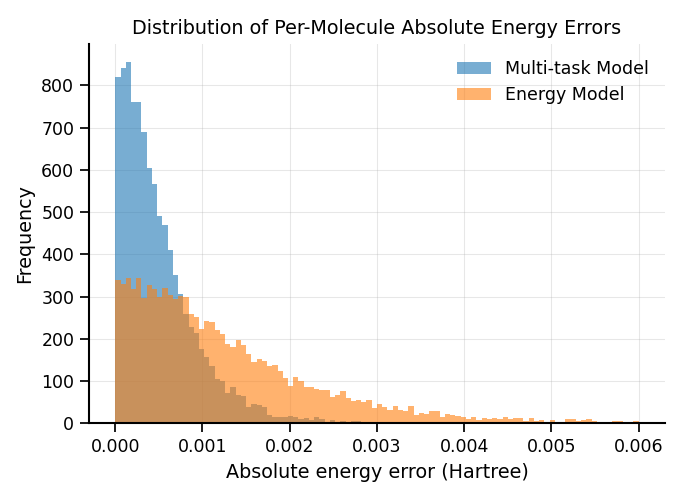}
    \caption{Distribution of per-molecule absolute energy errors on the QM9 test set for the multi-task and energy-only models. The multi-task model exhibits a sharper concentration near zero and a substantially shorter high-error tail, indicating both lower typical prediction error and fewer large-error cases.}
    \label{fig:histograms}
\end{figure}

In~\Cref{tab:comparison}, we report the test-set energy MAE on QM9 as a function of training-set size for both energy-only and multi-task models. For the energy-only setting, we compare HotPP models with output channel widths of 64 and 100. This comparison is intended to assess whether increasing the representational capacity of the model improves energy prediction accuracy.

Before comparing the energy-only and multi-task settings, we first examine
the effect of channel width. The results show that the 64-channel model
consistently outperforms the 100-channel model across all training-set sizes.
The difference is substantial at small data scales, with MAEs of \(481.64\)
and \(655.79\)~meV for \(1{,}000\) training samples, and remains present at
the largest training-set size, where the corresponding MAEs are \(36.46\)
and \(53.74\)~meV. The complete comparison is shown
in~\Cref{fig:learningcurves}.

Thus, within the range considered here, increasing the channel width does
not improve energy accuracy and instead leads to systematically worse
generalization.

The wider model was also more difficult to optimize. In our experiments, stable training of the 100-channel model required a smaller learning rate than the 64-channel model. Because the input dimension of the first linear layer in the ($\phi$) network scales with the channel width, increasing the width also increases the number of trainable parameters and the dimensionality of the optimization problem. Previous work has shown that the optimization configuration associated with the best generalization performance can depend systematically on network width and parameterization~\cite{park2019effect}. In particular, Park et al.\ identify the effective SGD noise scale, which depends on quantities such as the learning rate and batch size, as an important factor governing width-dependent generalization behavior.

We therefore hypothesize that the optimization configuration used for the 100-channel model may not have produced an effective SGD noise scale close to its optimum. This could partly explain why the greater nominal capacity of the wider model did not translate into improved test accuracy. However, because we did not perform a systematic sweep over learning rates and batch sizes for each channel width, this interpretation remains a hypothesis rather than a confirmed explanation. The 100-channel model may require more extensive width-specific hyperparameter tuning or regularization. Overall, the results indicate that the observed performance difference should not be interpreted solely in terms of model expressivity: for the present task and optimization setup, the 64-channel configuration provides a more favorable balance between capacity, optimization stability, and generalization.

We next compare the energy-only model with the multi-task models trained using different fractions of the available density supervision. The performance of the models can be seen in ~\Cref{fig:learningcurves}. The results show a substantial improvement from introducing the auxiliary density objective. The energy-only model achieves a test MAE of 36.46~meV at the largest training-set size, whereas the multi-task model trained with the full density dataset reaches 14.48~meV.

The improvement provided by density supervision changes markedly with training-set size. At \(1\mathrm{k}\) and \(10\mathrm{k}\) training samples, increasing the fraction of density-labelled structures produces large gains: the full-density model clearly outperforms the \(10\%\)- and \(1\%\)-density variants. However, this gap narrows as the number of energy-labelled training structures increases. At \(50\mathrm{k}\) samples, the full-density and \(10\%\)-density models achieve similar MAEs of \(19.37\) and \(22.34\)~meV, respectively, while at \(123\mathrm{k}\) samples the difference is reduced further to only \(0.35\)~meV. Even the \(1\%\)-density model reaches \(18.75\)~meV at the largest training-set size, substantially outperforming the energy-only baseline. These results indicate diminishing returns from increasing density coverage as the energy training set grows: dense supervision is most valuable in the low-data regime, whereas at larger data scales a comparatively small density-labelled subset is sufficient to recover most of the benefit of full density supervision.

We also compare the training dynamics of the best-performing multi-task model, trained with full density supervision, and the best-performing energy-only model with a channel width of 64. As shown in~\Cref{fig:trainingcurves}, the multi-task model reaches a lower training energy error earlier and maintains an advantage throughout most of training. The difference is most pronounced during the early and intermediate stages, indicating faster convergence under density supervision. Although the training curves become relatively close toward the end of training, the multi-task model achieves a substantially lower test-set MAE. This suggests that the main benefit of density supervision is not merely a reduction in training error, but improved generalization to unseen molecules. Finally, we examine the distribution of per-molecule absolute errors on the QM9 test set, as shown in~\Cref{fig:histograms}. The multi-task model exhibits a substantially sharper concentration near zero, whereas the energy-only model has a broader distribution with a longer high-error tail. This indicates that the improvement in mean test MAE is not driven by only a small subset of molecules. Rather, density supervision reduces the prediction error across a broad portion of the test set and also decreases the frequency of large-error cases. The narrower error distribution therefore provides further evidence that the multi-task model generalizes more consistently than the energy-only baseline.

These results indicate that density supervision provides a strong inductive bias for energy prediction. The density objective does not merely introduce an additional output target; it constrains the shared encoder to represent electronic-structure information that is physically relevant to the total energy. As discussed above, the energy and density heads operate on the same latent representation. Gradients from the density objective therefore shape the encoder toward features that must support the reconstruction of a spatially resolved electronic quantity, rather than only a single scalar target. The resulting representation appears to be more informative and more transferable to the energy-prediction task.

The performance as a function of density coverage also suggests diminishing returns from additional density labels. Increasing the density fraction from 1\(\%\) to 10\(\%\) produces a further improvement, but increasing it from \(10\%\) to the full density dataset yields only a comparatively small gain at larger training set sizes. Thus, the benefit of density supervision appears to approach saturation well before density labels are available for every training structure. In particular, the \(10\%\)-density model performs nearly as well as the fully supervised multi-task model, while the \(1\%\)-density model still substantially outperforms the energy-only baseline.

One possible interpretation is that electron-density supervision provides a particularly information-rich training signal. A scalar energy label imposes only one constraint per molecular structure, whereas the density target contains spatially resolved information about the electronic distribution throughout the molecule. Even when density labels are available for only a small subset of structures, gradients from this auxiliary objective may be sufficient to steer the shared encoder toward a physically meaningful region of representation space. Once this representation has been established, the much larger set of energy labels can refine the energy predictor without requiring density supervision for every sample.

This interpretation is consistent with the frozen-representation probe, where the encoder trained with density supervision produced substantially more accurate downstream energy predictions than the energy-only encoder under an identical probe architecture. Taken together, the probe and the final multi-task results support the view that density supervision improves the structure of the learned latent representation itself. The principal benefit therefore appears to arise not only from joint optimization of two prediction heads, but from learning a representation that captures electronic information relevant to energy generalization.

These conclusions should nevertheless be taken cautiously. The experiments demonstrate that sparse density supervision is sufficient to produce large improvements in energy test MAE, but they do not directly identify which components of the learned representation are responsible for the gain. Moreover, the precise point at which performance saturates remains uncertain because only a limited set of density-label fractions was evaluated and variability across independent runs was not quantified. Subject to these limitations, the results suggest that density labels can be used selectively: a relatively small density-labelled subset may recover most of the benefit of full multi-task supervision while substantially reducing the cost of generating and storing density data.
\paragraph{Out-of-distribution test.}
To investigate whether the benefits of density supervision transfer to out-of-distribution molecular systems, we evaluate the models on the Alchemy dataset~\cite{chen2019alchemy}. For this comparison, we use the best-performing energy-only configuration, with a channel width of 64, and the best-performing multi-task model, trained with full density supervision. The 100-channel energy-only model is excluded from the primary OOD comparison because it performs consistently worse on the QM9 test set across all training-set sizes. This ensures that the multi-task model is compared against the strongest available energy-only baseline rather than against a weaker configuration.

\FloatBarrier

\begin{table}[t]
\centering
\caption{Calibrated test-set MAE of formation-energy predictions on the Alchemy subsets. MAE values are reported in meV.}
\label{tab:alchemy_mae}
\setlength{\tabcolsep}{4pt}
\begin{tabular}{lccc}
\toprule
\textbf{Subset}
& \shortstack{\textbf{Energy-only}\\\textbf{MAE}}
& \shortstack{\textbf{Multi-task}\\\textbf{MAE}}
& \shortstack{\textbf{Relative}\\\textbf{improvement}} \\
\midrule
atom\_9  & 604.09 & 549.67 & 9.0\% \\
atom\_10 & 259.87 & 241.91 & 6.9\% \\
atom\_11 & 371.71 & 309.39 & 16.8\% \\
atom\_12 & 517.02 & 371.71 & 28.1\% \\
\bottomrule
\end{tabular}
\end{table}
\begin{figure}[t!]
    \centering
    \includegraphics[width=0.99\linewidth]{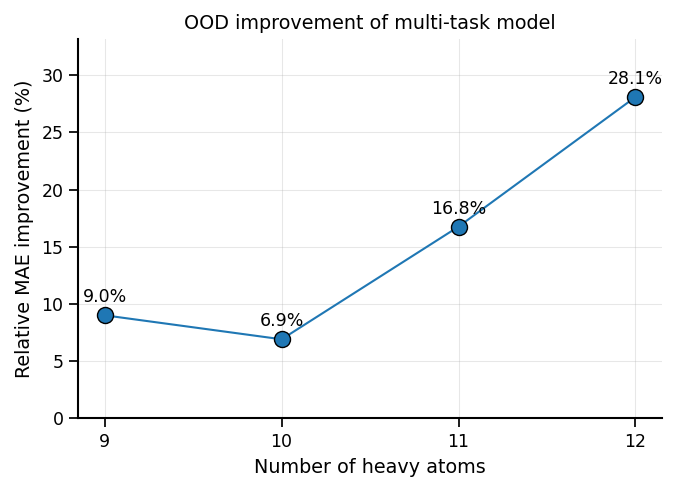}
    \caption{Relative MAE improvement of the multi-task model over the energy-only model trained on QM9 molecules when extrapolating to Alchemy subsets with different numbers of heavy atoms. Positive values indicate lower calibrated MAE for the multi-task model. The improvement increases for larger molecules, suggesting that the multi-task representation transfers more strongly to increasingly out-of-distribution molecular sizes.}
    \label{fig:alchemy_improvement}
\end{figure}

Because the ELECTRA models were trained on formation energies, we construct corresponding Alchemy targets using the same atomic reference energies as in the QM9 experiments. This places the two datasets on a common target definition, but does not remove the distribution shift between them. In particular, QM9 contains molecules with at most 9 heavy atoms, whereas Alchemy contains molecules with 9--12 heavy atoms and exhibits a broader range of molecular sizes and compositions. Consequently, systematic prediction errors can accumulate with molecular size and produce subset-dependent offsets and scale differences. The raw formation energy test dataset MAE may therefore reflect both transfer quality and dataset-level calibration mismatch.

To separate these effects, we apply a linear calibration as a diagnostic correction before computing the MAE. For each model and evaluated Alchemy subset, we fit
\begin{equation}
    \hat{E}^{\mathrm{cal}}_i
    =
    a\hat{E}_i+b,
\end{equation}
where \(\hat{E}_i\) is the original model prediction and \(a\) and \(b\) are obtained by least-squares fitting to the Alchemy target energies. The calibrated MAE is then
\begin{equation}
    \mathrm{MAE}_{\mathrm{cal}}
    =
    \frac{1}{N}
    \sum_{i=1}^{N}
    \left|
    E_i-\hat{E}^{\mathrm{cal}}_i
    \right|,
\end{equation}
where \(E_i\) is the target formation energy. This calibrated metric is used as a diagnostic measure of how well each model preserves relative energy variation after correcting for dataset-level offset and scale shifts. Given that Alchemy is organized by heavy-atom count and model performance varies strongly with molecular size, we apply the calibration separately within the atom\_9, atom\_10, atom\_11, and atom\_12 subsets. We restrict the evaluation to molecules that do not contain sulfur or chlorine, since these elements are absent from QM9. The resulting calibrated test-set MAEs of the formation-energy predictions are reported in~\Cref{tab:alchemy_mae}, while the relative improvements are shown in~\Cref{fig:alchemy_improvement}.

The multi-task model achieves a lower calibrated formation-energy MAE than the energy-only model for every heavy-atom-count subset. For atom\_9 molecules, the MAE decreases from \(604.09\) to \(549.67\)~meV, corresponding to a relative improvement of \(9.0\%\). For atom\_10, the MAE decreases from \(259.87\) to \(241.91\)~meV, an improvement of \(6.9\%\). The benefit becomes larger for the more strongly out-of-distribution subsets: the atom\_11 MAE decreases from \(371.71\) to \(309.39\)~meV, while the atom\_12 MAE decreases from \(517.02\) to \(371.71\)~meV, corresponding to improvements of \(16.8\%\) and \(28.1\%\), respectively. This trend suggests that the representation learned with density supervision transfers more effectively as molecular size departs further from the QM9 training distribution. Nevertheless, the calibrated Alchemy errors remain substantially larger than the corresponding in-distribution QM9 errors, indicating that neither model retains its original level of accuracy under this molecular-size shift.

%% file: sec/7_conclusions.tex
\section{Conclusion}

In this work, we investigated whether electron-density supervision can improve molecular energy prediction by serving as a physically meaningful proxy for electronic-structure learning. Starting from the ELECTRA~\cite{electra} density-prediction framework, we introduced an energy readout and compared an energy-only model with a multi-task model trained jointly on electron density and molecular energy.

Density supervision consistently improves energy prediction, remains effective even when applied to only a small fraction of the training data, and yields lower calibrated errors in our out-of-distribution evaluation than energy-only training.

The results support the hypothesis that electron density provides useful auxiliary supervision for molecular energy prediction. Unlike energy, which is a global scalar label, the density provides spatially resolved information about the electronic structure, encouraging the shared representation to encode features related to bonding, charge distribution, and local chemical environments. In this sense, density supervision acts as a physically grounded proxy task that helps shape representations useful for predicting quantum-mechanical observables.

Several limitations remain. First, the optimal architectural configuration is not necessarily the same for energy-only and multi-task learning. The density objective places different demands on the shared representation than scalar energy prediction, and architectural choices such as the HotPP channel width therefore cannot be assumed to affect the two objectives in the same way. Moreover, even within a fixed objective, increasing representation width does not necessarily lead to improved performance; we observe that changes in channel width can produce non-monotonic and sometimes unexpected effects on energy accuracy.

However, the frozen-representation probe controls for the channel width confound by comparing encoders with identical architectures that directly match the multi-task implementation. Under this controlled setting, the density-supervised encoder still produces substantially better downstream energy predictions than the energy-only encoder. This indicates that the observed improvement cannot be attributed solely to greater model expressivity, but is also consistent with density supervision producing a more energy-relevant learned representation.

Second, joint density--energy training increases computational cost because it requires optimizing a high-dimensional spatial field in addition to the scalar energy. Future work could address this by sampling the density grids during training to only use a small number of grid points, potentially reducing the training costs to a fraction.

Third, although density supervision substantially improves the energy predictions within our experimental setup, the best-performing models remain below the strongest results reported for QM9 energy prediction in recent benchmark studies~\cite{qm9benchmark}. This indicates that the present contribution should be interpreted primarily as evidence for the benefit of density-guided representation learning, rather than as a new state-of-the-art result on QM9.

Finally, the out-of-distribution evaluation is limited to the Alchemy dataset, so broader testing on additional chemical spaces and larger molecular systems is needed to establish generality.

Overall, this work shows that electron-density supervision improves the accuracy, data efficiency, convergence behavior, and out-of-distribution transfer of molecular energy models. The frozen-representation probe further indicates that these gains arise from a more energy-relevant shared representation rather than solely from increased model capacity. Moreover, most of the improvement is retained with sparse density supervision, suggesting that full density coverage is not required. Future work should quantify the variability across independent runs, determine more precisely how performance saturates with density coverage, investigate adaptive task weighting and gradient interactions, and test whether the observed benefits extend to broader chemical spaces, larger molecular systems, and other model architectures.

%% file: sec/8_acknowledgements.tex
\section*{Acknowledgements}

The authors acknowledge the computational resources provided by the Niflheim high-performance computing cluster at the Technical University of Denmark, as well as the financial support from the Independent Research Foundation Denmark with grant no. 3164-00297B (ADANA), from the Novo Nordisk Foundation with grant number NNF25OC0101622 (AutoMLP), and grant no. 2035-00232B (TeraBatt).

%% file: sec/9_code.tex
\section*{Code availability}

The code used to train and evaluate the models in this work is publicly available at \url{https://github.com/vdmnsaiioa/ELECTRA_mm}.

%% file: sec/appendix.tex
\section{Hyperparameters}

\begin{table*}[t]
    \centering
    \caption{Hyperparameters for the energy-only and multi-task models. Shared hyperparameters are listed once, while model-specific hyperparameters are shown separately.}
    \label{tab:model_hyperparameters}
    \setlength{\tabcolsep}{10pt}
    \begin{tabular}{lccc}
        \toprule
        \multicolumn{4}{l}{\textbf{Shared hyperparameters}} \\
        \midrule
        \textbf{Hyperparameter} & \multicolumn{3}{c}{\textbf{Value}} \\
        \midrule
        Graph radius cutoff
        & \multicolumn{3}{c}{$8.0~\text{\AA}$} \\
        HotPP body layers
        & \multicolumn{3}{c}{$6$} \\
        Gaussians per electron $(M_e)$
        & \multicolumn{3}{c}{$90$} \\
        GNN $L_{\max}$ -- body layers
        & \multicolumn{3}{c}{$3$} \\
        GNN $L_{\max}$ -- head layers
        & \multicolumn{3}{c}{$3$} \\
        Body order $(N_{\max})$
        & \multicolumn{3}{c}{$3$} \\
        Precision
        & \multicolumn{3}{c}{Float} \\
        Optimizer
        & \multicolumn{3}{c}{Adam~\cite{kingma2014adam}} \\
        Weight decay
        & \multicolumn{3}{c}{$0$} \\
        \midrule
        \multicolumn{4}{l}{\textbf{Model-specific hyperparameters}} \\
        \midrule
        \textbf{Hyperparameter}
        & \textbf{Energy-only 64 ch.}
        & \textbf{Energy-only 100 ch.}
        & \textbf{Multi-task} \\
        \midrule
        Graph network channel width
        & $64$
        & $100$
        & $750$ \\
        Energy channel width
        & --
        & --
        & $100$ \\
        Initial learning rate $LR_{\text{initial}}$
        & $4.0 \times 10^{-5}$
        & $4.0 \times 10^{-6}$
        & $4.0 \times 10^{-6}$ \\
        Fine-tuning learning rate $LR_{\text{fine}}$
& $5.0 \times 10^{-7}$
& $1.0 \times 10^{-7}$
& $1.0 \times 10^{-7}$ \\
        Energy-loss weight $\alpha$
        & --
        & --
        & $10$ \\
        Density-loss weight $\beta$
        & --
        & --
        & $1$\\
        \bottomrule
    \end{tabular}
\end{table*}

The multi-task model requires a substantially larger GNN channel width to provide sufficient representational capacity for learning the spatial electron-density target in addition to the scalar energy. In preliminary experiments, using the full \(750\)-channel representation directly for energy prediction resulted in less stable optimization and poorer energy convergence. The multi-task architecture therefore projects the HotPP features to a lower-dimensional energy representation using a fully connected neural network before the final UGP energy readout. This reduced dimensionality is reported in~\Cref{tab:model_hyperparameters} as the energy channel width.

The energy-only models do not use this additional projection, since their HotPP channel widths are selected directly for the scalar energy-prediction task. We evaluate energy-only models with channel widths of \(64\) and \(100\), giving corresponding UGP input dimensions of \(64\) and \(100\). The multi-task model also uses an energy-channel width of \(100\) after projection, although its underlying GNN representation has a width of \(750\). The \(64\)-channel energy-only model uses a larger learning rate, while the \(100\)-channel energy-only and multi-task models use the same smaller learning rate to obtain stable training.

Training was performed in two phases. Each model was first optimized using the initial learning rate reported in~\Cref{tab:model_hyperparameters}. The selected checkpoint was then further optimized using a reduced fine-tuning learning rate. The \(64\)-channel energy-only model used a fine-tuning learning rate of \(5.0\times10^{-7}\), while the \(100\)-channel energy-only and multi-task models used \(1.0\times10^{-7}\).

All multi-task variants, including the models trained with \(1\%\), \(10\%\), and full density supervision, use the same architecture and optimization hyperparameters. They differ only in the fraction of training structures for which the density loss is applied. The reported configurations were selected to provide stable training and strong validation performance for each objective, rather than to enforce identical internal widths across models with different prediction tasks.

\section{Atomic reference energies}
\label{app:atomic_references}

To remove the dominant composition-dependent contribution from the total molecular energies, we define element-specific atomic reference energies for
\(\mathrm{H}\), \(\mathrm{C}\), \(\mathrm{N}\), \(\mathrm{O}\), and \(\mathrm{F}\).
These values are obtained from the total energies of a small set of reference molecules:
\(\mathrm{CH_3OH}\), \(\mathrm{CH_4}\), \(\mathrm{H_2O}\), \(\mathrm{NH_3}\), and \(\mathrm{CF_4}\).

The corresponding molecular energies are
\[
\begin{aligned}
E(\mathrm{CH_3OH}) &= -115.679136, \\
E(\mathrm{CH_4})   &= -40.478930, \\
E(\mathrm{H_2O})   &= -76.404702, \\
E(\mathrm{NH_3})   &= -56.525887, \\
E(\mathrm{CF_4})   &= -437.484875.
\end{aligned}
\]

The oxygen reference energy is first obtained from the energy difference between methanol and methane,
\[
E_{\mathrm{O}}^{\mathrm{ref}}
=
E(\mathrm{CH_3OH})
-
E(\mathrm{CH_4}),
\]
which gives
\[
E_{\mathrm{O}}^{\mathrm{ref}}
=
-75.200206.
\]

The hydrogen reference energy is then determined from the water molecule,
\[
E_{\mathrm{H}}^{\mathrm{ref}}
=
\frac{
E(\mathrm{H_2O})
-
E_{\mathrm{O}}^{\mathrm{ref}}
}{2},
\]
yielding
\[
E_{\mathrm{H}}^{\mathrm{ref}}
=
-0.602248.
\]

The carbon reference energy is obtained from methane,
\[
E_{\mathrm{C}}^{\mathrm{ref}}
=
E(\mathrm{CH_4})
-
4E_{\mathrm{H}}^{\mathrm{ref}},
\]
which gives
\[
E_{\mathrm{C}}^{\mathrm{ref}}
=
-38.069938.
\]

Similarly, the nitrogen reference energy is obtained from ammonia,
\[
E_{\mathrm{N}}^{\mathrm{ref}}
=
E(\mathrm{NH_3})
-
3E_{\mathrm{H}}^{\mathrm{ref}},
\]
yielding
\[
E_{\mathrm{N}}^{\mathrm{ref}}
=
-54.719143.
\]

Finally, the fluorine reference energy is obtained from carbon tetrafluoride,
\[
E_{\mathrm{F}}^{\mathrm{ref}}
=
\frac{
E(\mathrm{CF_4})
-
E_{\mathrm{C}}^{\mathrm{ref}}
}{4},
\]
which gives
\[
E_{\mathrm{F}}^{\mathrm{ref}}
=
-99.853734.
\]

The resulting atomic reference energies are summarized in
\Cref{tab:atomic_reference_energies}.

\begin{table}[t]
\centering
\caption{Atomic reference energies used to construct the composition-corrected molecular energy targets.}
\label{tab:atomic_reference_energies}
\begin{tabular}{lc}
\toprule
\textbf{Element}
& \textbf{Reference energy} \\
\midrule
\(\mathrm{H}\) & \(-0.602248\) \\
\(\mathrm{C}\) & \(-38.069938\) \\
\(\mathrm{N}\) & \(-54.719143\) \\
\(\mathrm{O}\) & \(-75.200206\) \\
\(\mathrm{F}\) & \(-99.853734\) \\
\bottomrule
\end{tabular}
\end{table}

For a molecule with total energy \(E_{\mathrm{tot}}\), the target energy used for training is
\[
E_{\mathrm{form}}
=
E_{\mathrm{tot}}
-
\sum_{a \in \{\mathrm{H},\mathrm{C},\mathrm{N},\mathrm{O},\mathrm{F}\}}
n_a E_a^{\mathrm{ref}},
\]
where \(n_a\) denotes the number of atoms of element \(a\) in the molecule.